\documentclass[11pt,a4paper]{article}
\usepackage{amsmath}
\usepackage{amssymb}
\usepackage{graphicx}
\usepackage{subcaption}
\usepackage{bbm}
\usepackage{bm}
\usepackage{breqn}
\usepackage{enumitem}
\usepackage[compat=1.1.0]{tikz-feynman}

\newcommand{\beq}{\begin{equation}}
\newcommand{\eeq}{\end{equation}}
\newcommand{\bea}{\begin{eqnarray}}
\newcommand{\eea}{\end{eqnarray}}                  
\newcommand{\bal}{\begin{align}}
\newcommand{\eal}{\end{align}}
\newcommand{\bpm}{\begin{pmatrix}}
\newcommand{\epm}{\end{pmatrix}}
\newcommand{\Ds}{\,{\slash \kern -8pt D}}

\tikzfeynmanset{warn luatex=false}

\makeatletter
\let\cat@comma@active\@empty
\makeatother

\begin{document}

\title{\textbf{Fixed points of the CKM matrix renormalization group running to all orders in perturbation theory}}

\author{Brian P. Dolan\footnote{email: bdolan@stp.dias.ie} \\
  \textit{\small School of Theoretical  Physics}\\
  \textit{\small Dublin Institute for Advanced Studies}\\
  \textit{\small 10 Burlington Rd., Dublin, Ireland}\\
  and \\
\textit{\small Department of Physics, Maynooth University}\\
  \textit{\small Maynooth, Co.~Kildare, Ireland}\\
}

  \maketitle
\begin{abstract}  
    A proof is given that the six fixed points of the massless 1-loop renormalization group running of the CKM matrix in the Standard Model  found previously at 1-loop are fixed points to all orders in perturbation theory.  A previous proof required an extra assumption on the running which is eliminated in the proof presented here.

  \rightline{DIAS-STP-26-19}
  \end{abstract}

\section{Introduction}

The running of the couplings in the Standard Model is  central to particle physics,
indeed all of quantum field theory,  and 1-loop results for the gauge couplings are well known to all students of modern physics.   While the running of the Yukawa couplings and the Higgs 4-point coupling are important for understanding the stability of the Higgs vacuum,
\cite{Stability, Shaposhnikov}, the running of the CKM parameters is not so well known: the reasons for this are probably two-fold: firstly even at 1-loop the running is rather complicated, and secondly the observed Yukawa couplings are so small that the running of the CKM parameters is of little physical significance.  Nevertheless it is conceptually important and a full understanding of the running of mixing matrices is likely to be essential for applications beyond the Standard Model, such as models involving right-handed neutrinos or models with more exotic matter content.
Partial results for the Yukawa sector, including the CKM parameters, were obtained in 
\cite{M+P} and  \cite{B+G}, the full 1-loop running being  given in
\cite{Babu} and \cite{Denner+Sack}; some 2-loop calculations were presented in
\cite{M+V} and \cite{L+X} and 3-loop  results in \cite{3-loops, HMS}.

The 1-loop running was revisited in  \cite{AEHNPS} where it was shown that there are six fixed points $V^*_\alpha$, $\alpha=1,\cdots,6$, of the running of the CKM matrix $V\in U(3)$, and these six matrices form a unitary representation of the symmetric group $S_3$ acting on  three objects.  
 An argument was given in \cite{CKM-running} that these special six points  are fixed points of the CKM matrix, not only at all orders of perturbation theory, but even non-perturbatively.\footnote{I thank Charles Nash for discussions on this point.}  The proof relies on the fact that the space of parameters of the CKM is matrix is a double coset which is
\beq U(1) \times U(1) \backslash SU(3) / U(1) \times U(1),\label{double-coset}\eeq
the left-action of the $SU(3)$ Cartan sub-algebra, $U(1)\times U(1)$, on the flag manifold 
\[F_3=SU(3) / U(1)\times U(1).\]
The $U(1)$'s are quark phases and $U(1)\times U(1) =T \subset  SU(3)$ is the Cartan torus of $SU(3)$, acting on the left and right of the fundamental representation of $SU(3)$ in \eqref{double-coset}.\footnote{This double coset is topologically $S^4$, \cite{Buchstaber+Terzic}, albeit a non-differentiable $S^4$ (similar to the way the surface of a cube is topologically $S^2$, but it is not differentiable at the  vertices and across the edges).  I thank C.~Nash for pointing reference \cite{Buchstaber+Terzic} 
out to me.} 

The proof  in \cite{CKM-running} that 
$V^*_\alpha$ are fixed points of the CKM matrix non-perturbatively assumes that the RG running on the double coset can be lifted to an RG running on $F_3$, and that the left action of the Cartan sub-algebra (CSA) on $F_3$ commutes with the RG flow on $F_3$. The purpose of the present work is to present an alternative proof  that the six points $V^*_\alpha$ are fixed points of the RG to all orders in perturbation theory that does not use this assumption. 
The 1-loop $\beta$-functions for the CKM matrix are reviewed in \S \ref{sec:1-loop} and \S \ref{sec:all-loops} gives the main argument.

\section{The CKM matrix fixed points at 1-loop \label{sec:1-loop}}

The Yukawa couplings in the Standard Model are
\beq  {\cal L}_{Yukawa} = - \sum_{\bar a, b =1}^3 \left( Y_{\bar a b} \overline \Psi_L^{\bar a} \Phi_c \Psi^b_R
+ Y'_{\bar a b}\overline \Psi_L^{\bar a}\Phi\Psi^{\prime b}_R
+ Y'_{l,\bar a b}\overline \Psi_{l,L}^{\bar a}\Phi\Psi^{\prime b}_{l,R}\right)
\ - \ \mbox{h.c.}
\eeq
where 
$\Phi=\bpm \phi^+\\  \phi^0 \epm $ is the Higgs doublet;
$\Phi_c=-i \sigma_2\Phi^*$ the conjugate Higgs;
$\Psi_L^a$ and $\Psi_{l,L}^a$  are left-handed $SU(2)$ doublets; $\Psi^a_R$,  $\Psi^{\prime a}_R$ and $\Psi^{\prime a}_{l,R}$
right-handed singlets:
\[ \Psi_L = \bpm u_{L} & c_L & t_L\\ d_L & s_L & b_L\epm, \quad
 \Psi_R = (u_R, c_R, t_R )\quad \mbox{and} \quad 
  \Psi'_R = (d_R,  s_R,  b_R ) . 
\]
and
\[ \Psi_{l,L} = \bpm \nu_{e,L} & \nu_{\mu,L} & \nu_{\tau,L}\\ e_L & \mu_L & \tau_L\epm, \quad
 \Psi'_{l,R} = (e_R, \nu_R, \mu_R ) . 
\]
The Yukawa couplings $Y_{\bar a b}$, $Y'_{\bar a b}$  and   $Y'_{l,\bar a b}$
are $3\times 3$ complex matrices, with $a,b=1,2,3$ labelling generations.

 Using dimensional regularisation, with $t=\ln \mu$ and $ \dot \  =\frac{d}{d t}$, the massless running is  \cite{Babu}, \begin{align}
\dot Y  &=
  \frac{1 }{16\pi^2}\left[\frac{3}{2}\left( Y Y^\dagger-   Y'Y^{\prime\dagger}\right)
                                   +4 \pi (\Pi -\alpha){\bm 1}\right]Y,
  \\
 \dot Y'  &=
  \frac{1 }{16\pi^2}\left[\frac{3}{2}\left( Y' Y^{\prime \dagger}-   Y Y^\dagger\right)
                       + 4 \pi (\Pi -\alpha'){\bm 1}\right]Y',
 \\
   \dot  Y'_l  &=
  \frac{1 }{16\pi^2}\left[\frac{3}{2}Y' Y^{\prime \dagger}
                       + 4 \pi (\Pi -\alpha'_l){\bm 1}\right]Y'.
 \end{align}
           $\Pi $ arises from summing over fermion loops in the Higgs leg of the Yukawa vertex and $\alpha$ and $\alpha'$ are quadratic combinations of  the  gauge
couplings $\alpha_3$, $\alpha_2$ and $\alpha_1$: explicitly
\[\Pi  =  \frac{1}{4 \pi}Tr\Bigl\{ 3(Y Y^\dagger+Y' Y^{\prime\dagger})+Y_l' Y_l^{\prime\dagger}\Bigr\}\] 
and
\[\alpha = 8\alpha_3 +\frac{9}{4}\alpha_2 + \frac{17}{12}\alpha_1, \qquad 
\alpha'= 8\alpha_3 +\frac{9}{4}\alpha_2 + \frac{5}{12}\alpha_1,\qquad
\alpha'_l=\frac{3}{4} (3 \alpha_2 + 5 \alpha_1),\]
$\alpha_i = \frac{g_i^2}{4\pi}$ being the gauge couplings.

 In terms of Hermitian matrices 
\[ Z=\frac{1}{4\pi} Y Y^\dagger,\quad   Z' = \frac{1} {4\pi}  Y' Y^{\prime \dagger}
\quad \mbox{and}  \quad Z_l' = \frac{1} {4\pi}  Y_l' Y_l^{\prime \dagger},\]
    the RG evolution is
              \begin{align}
  \dot  Z &=
\frac{1}{4\pi} \left\{   3 Z^2-\frac{3}{2}  \left(Z Z' + Z' Z\right)
 +2(\Pi  -\alpha) Z\right\},\label{eq:dZdt}\\
  \dot  Z' &=
\frac{1}{4 \pi} \left\{   3 Z^{\prime 2}-\frac{3}{2}  \left(Z Z' + Z' Z\right)
                           +2(\Pi  -\alpha')Z'\right\} ,\label{eq:dZ'dt}\\
                             \dot  Z_l' &=
\frac{1}{4 \pi} \left\{   3 Z_l^{\prime 2}
                     +2(\Pi  -\alpha'_l)Z_l'\right\} .\label{eq:dZ_l'dt}
 \end{align}

   The three  matrices $Z$, $Z'$ and $Z_l'$ can be  diagonalized with $U$, $U'$ and $U_l'\in SU(3)$:
\beq \Lambda  = U^\dagger Z U,
  \qquad \Lambda ' = U^{\prime \dagger} Z' U' \qquad \mbox{and} \qquad  \Lambda_l ' = U_l^{\prime \dagger} Z_l' U_l' 
  \label{eq:ZLambda}\eeq
  where the diagonal components of $\Lambda$, $\Lambda'$  and $\Lambda_l'$ are related to the usual Yukawa couplings by
 \[\Lambda _a=\frac{y_a^2}{4\pi} ,  \qquad 
    \Lambda '_a= \frac{y_a^{\prime 2}}{4\pi} \qquad \mbox{and} \qquad    \Lambda '_{l,a}= \frac{y_{l,a}^{\prime 2}}{4\pi}.\]
        There is no physics in $U_l'$: $Z_l'$ can simply be diagonalized to give the running of the three leptonic Yukawa couplings immediately,\footnote{Three sterile right-handed neutrinos require introducing the PMNS matrix, and the analysis presented here for the CKM matrix  would go through the same way for the PMNS matrix.} but $U$ and $U'$ generate the CKM matrix
\beq V=U^\dagger U'.\label{eq:CKM}\eeq
 
In terms of $\Lambda$ and $\Lambda'$ equations \eqref{eq:dZdt} and  \eqref{eq:dZ'dt}
read
\begin{align}
  \kern -30pt   \dot \Lambda- i[A,\Lambda]&=
 \frac{1}{4\pi}\left\{3\Lambda^2 -
 \frac{3}{2}[\Lambda V \Lambda' V^\dagger +  V \Lambda' V^\dagger\Lambda ]
 +2\bigl( \Pi   -\alpha\bigr) \Lambda,\right\}\label{eq:Lambda}\\
   \dot\Lambda'- i[A',\Lambda']&=
     \frac{1}{4 \pi} \left\{   3 \Lambda^{\prime 2}
    - \frac{3}{2}[\Lambda'  V^\dagger \Lambda  V +  V^\dagger \Lambda  V \Lambda']
                                              +2\bigl( \Pi  -\alpha'\bigr) \Lambda' \right\},\label{eq:Lambda'}
\end{align}
with
\[A = i U^\dagger \dot U, \quad \mbox{and} \quad A' = i U^{\prime \dagger}  \dot U'.\]

The off-diagonal components of the matrices  $A$, $A'$,
$\Lambda V \Lambda' V^\dagger +  V \Lambda' V^\dagger\Lambda$ and
$\Lambda'  V^\dagger \Lambda  V +  V^\dagger \Lambda  V \Lambda'$,
determine the running of the CKM parameters, \cite{CKM-running}. 

The 1-loop RG running of  the Yukawa couplings $y_a$ and $y_a'$ themselves 
is 
\begin{align}
  \dot \Lambda&=
 \frac{1}{4\pi}\left\{3\Lambda^2 -
 \frac{3}{2}[\Lambda V \Lambda' V^\dagger +  V \Lambda' V^\dagger\Lambda ]_{diag}
 +2\bigl( \Pi   -\alpha\bigr) \Lambda,\right\}\label{eq:A}\\
\dot\Lambda'&=
     \frac{1}{4 \pi} \left\{   3 \Lambda^{\prime 2}
    - \frac{3}{2}[\Lambda'  V^\dagger \Lambda  V +  V^\dagger \Lambda  V \Lambda']_{diag}
                      +2\bigl( \Pi  -\alpha'\bigr) \Lambda' \right\},\label{eq:A'}
\end{align}
where $[\cdots]_{diag}$ denotes the diagonal component of the matrix.

 We use the  standard parameterisation of the CKM matrix,  
\begin{align}
  V&=\bpm
c_{3}c_{2} & s_{3}c_{2} & s_{2} e^{-i\delta} \cr
-s_{3}c_{1}-c_{3}s_{1}s_{2}e^{i\delta} 
& c_{3}c_{1} -s_{3}s_{1}s_{2}e^{i\delta} & s_{1}c_{2} \cr 
s_{3}s_{1} - c_{3}c_{1}s_{2}e^{i\delta} &
    -c_{3}s_{1}-s_{3}c_{1}s_{2}e^{i\delta} & c_{1}c_{2}\epm \hskip -2pt ,
      \label{eq:Vdelta}
\end{align}
where $c_i=\cos\theta_i$ and $s_i=\sin\theta_i$, $\theta_i$ being the mixing angles $\theta_1=\theta_{23}$, \textit{etc}.
The massless 1-loop $\beta$-functions for the CKM matrix vanish when
          \begin{alignat*}{4}
            \left(\theta_{1}, \theta_{2}, \theta_{3}\right)
= &\Bigl( 0,0,0\Bigr), \
      &   &   \left(0,0,\frac{\pi}{2}\right),\
&            &\left(\frac{\pi}{2},0,0\right),\
   &            &     \left(\frac{\pi}{2},0,\frac{\pi}{2}\right), 
   \\
  &
            \left(0,\frac{\pi}{2},0\right),\
         &       &     \left(\frac{\pi}{2},\frac{\pi}{2},0\right),\
            &   & \left(0,\frac{\pi}{2},\frac{\pi}{2}\right),\
   &              &\left(\frac{\pi}{2},\frac{\pi}{2},\frac{\pi}{2}\right),
          \end{alignat*}
          with $\delta=0$ or $\pi$.
Up to quark re-phasing,
only six of these sixteen fixed points are physically distinct:
extending the CKM matrix to  lie in  $U(3)$, rather than restricting it to $SU(3)$,
these six points are  \cite{AEHNPS}
{\scriptsize{\begin{alignat}{4}
    V_1^*=&\begin{pmatrix} 1&0&0\\ 0&1&0\\ 0 &0&1 \end{pmatrix},&
             \   & (\theta_{1}^*,\theta_{2}^*,\theta_{3}^*)=(0,0,0);&     \ 
 V^*_2=&\begin{pmatrix} 0&1&0\\ 1&0&0\\ 0 &0&1 \end{pmatrix},&
       \   &(\theta_{1}^*,\theta_{2}^*,\theta_{3}^*)=\Bigl(0,0,\frac{\pi}{2}\Bigr);\nonumber\\
      V_3^* =  &\begin{pmatrix} 0&1&0\\ 0&0&1\\ 1&0&0 \end{pmatrix},&
  \    &(\theta_{1}^*,\theta_{2}^*,\theta_{3}^*)= \Bigl(\frac{\pi}{2},0,\frac{\pi}{2}\Bigr);&  \ 
 V^*_4=&\begin{pmatrix} 1&0&0\\ 0&0&1\\ 0 &1&0  \end{pmatrix},&
     \  &(\theta_{1}^*,\theta_{2}^*,\theta_{3}^*)=\Bigl(\frac{\pi}{2},0,0\Bigr);\nonumber\\
 V_5^*=&  \begin{pmatrix} 0&0&1\\ 1  &0&0\\ 0 &1&0 \end{pmatrix},&
       \      &  (\theta_{1}^*,\theta_{2}^*,\theta_{3}^*)=
          \begin{cases}
            \Bigl(0,\frac{\pi}{2},\frac{\pi}{2}\Bigr),\\[0.5em]
            \Bigl(\frac{\pi}{2},\frac{\pi}{2},0\Bigr) ;
          \end{cases}&         \  
   V^*_6 =& \begin{pmatrix} 0&0&1\\ 0&1&0\\ 1 &0&0 \end{pmatrix},&
                  \  & (\theta_{1}^*,\theta_{2}^*,\theta_{3}^*)
                  =\begin{cases}
                   \Bigl(0,\frac{\pi}{2},0\Bigr),\\[0.5em]
               \Bigl(\frac{\pi}{2},\frac{\pi}{2},\frac{\pi}{2}\Bigr).
               \end{cases}\label{eq:S_3}
\end{alignat}
}}
\noindent These are fixed points of the CKM matrix regardless of the values of the Yukawa couplings: they are not necessarily fixed points of the whole Yukawa sector,
the Yukawa couplings $y_a$, $y'_a$ and $y'_{l,a}$ can still be running, but if the CKM matrix ever hits any of the fixed points in \eqref{eq:S_3} it remains there, \cite{CKM-running}.

These six matrices form a unitary representation of $S_3$, the symmetric group of three objects, which relates to the fact that this is the Weyl group of $SU(3)$,   \cite{AEHNPS} and \cite{CKM-running}.\footnote{In the $SU(3)$ representation of $V$       in \eqref{eq:Vdelta} some of the entries in \eqref{eq:S_3} are $-1$, but this is just a phase change of $\pi$ for some of the quarks which can be removed by lifting to $U(3)$.}
  
\section{The CKM matrix fixed points  at all loops\label{sec:all-loops}}

The $\beta$-functions at 2-loop were calculated in \cite{M+V}, and reproduced in \cite{L+X}: 
3-loop calculations were presented in \cite{3-loops,HMS}.
For our purposes the specific details are not important: we only need note that, with the external legs fixed all internal fermion lines must contract over family indices in a way that is consistent with matrix multiplication.
Any internal right-handed fermion propagator bounded by Yukawa vertices
will give contribute either $Z$ or $Z'$ to the associated Feynman diagram.

\bigskip

\hskip 50pt \hbox{\vtop
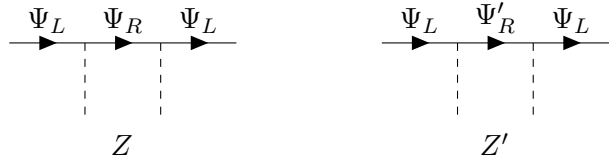
{\hbox{   \begin{tikzpicture}
    \begin{feynman}
      \vertex at (0,0)  (in);
   \vertex at (1,0)  (y1);
   \vertex at (2,0)  (y2);
   \vertex at (3,0)  (out);
    \vertex at (1,-1) (h1);
    \vertex at (2,-1) (h2);
      \diagram*{   
(in) --  [fermion, edge label =\(\Psi_L\)]  (y1), 
(y1) -- [fermion, edge label =\(\Psi_R\)] (y2),
 (y2) --[fermion, edge label =\(\Psi_L\)] (out) ,
    (y1) -- [scalar] (h1),
      (y2) -- [scalar] (h2),
};
    \end{feynman}
  \end{tikzpicture}}
\hskip 25pt \hbox{$Z$}}

\hskip -220pt

  \vtop{ \hbox{\begin{tikzpicture}
    \begin{feynman}
      \vertex at (0,0)  (in);
   \vertex at (1,0)  (y1);
   \vertex at (2,0)  (y2);
   \vertex at (3,0)  (out);
    \vertex at (1,-1) (h1);
    \vertex at (2,-1) (h2);
      \diagram*{   
(in) --  [fermion, edge label =\(\Psi_L\)]  (y1), 
(y1) -- [fermion, edge label =\(\Psi'_R\)] (y2),
 (y2) --[fermion, edge label =\(\Psi_L\)] (out) ,
    (y1) -- [scalar] (h1),
      (y2) -- [scalar] (h2),
};
    \end{feynman}
  \end{tikzpicture}}
\hskip 20pt \hbox{$Z'$}}
}
\bigskip

In general, at any order in perturbation theory, sub-diagrams can only generate linear combinations of matrix products such as
\[ \cdots Z^{q_1} Z^{\prime q_2} Z^{q_3} Z^{\prime q_4} Z^{q_5}\cdots \]
in the $\beta$-functions for $Z$ and $Z'$ and internal fermion loops are contracted over family indices, contributing traces of such combinations.
 For example the diagram

     
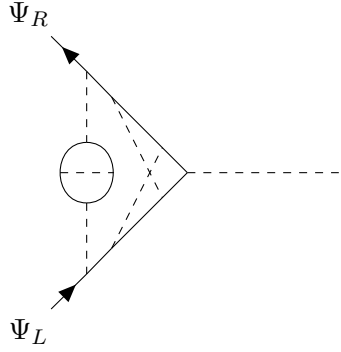
\centerline{
   \begin{tikzpicture}
    \begin{feynman}
        \vertex at (-2.1, 2.1) (fR)  {\(\Psi_R\)};   
        \vertex at (-2.1,-2.1) (fL)  {\(\Psi_L\)};
        \vertex at (0,0) (v) ;
        \vertex at (2,0) (h) ;
\vertex at (0.67-2,-1.34) (a); 
\vertex at (-1,-1) (b) ;
\vertex at (-1,1) (d) ;
\vertex at (1.67-2,0.33) (l) ;
\vertex at (1.67-2,-0.33) (m) ;
\vertex at (0.67-2,1.34) (c) ;
\vertex at (0.67-2,0.4) (e) ;
\vertex at (0.67-2,-0.4) (f) ;
\vertex at (0.32-2,0) (g) ;
\vertex at (1.03-2,0) (k) ;
                  \diagram*{
            (fL) --  [fermion] (a)  -- [plain] (b)  --[plain] (v) ,
            (v) -- [plain] (d) --[plain] (c) --[fermion] (fR),
           (c) -- [scalar] (e),
           (f) -- [scalar] (a),
           (e) -- [plain, half left] (f),
           (e) -- [plain, half right] (f),
             (g) -- [scalar] (k),
             (v) -- [scalar] (h),
             (b) -- [scalar] (l),
             (d) -- [scalar] (m),
                       };
                     \end{feynman}
                   \end{tikzpicture}
                 }
 \noindent will contribute terms with
\[ Z^3 Y, \ Z^2 Z' Y, \ Z Z' Z Y,\ Z' Z^2 Y,\ Z^{\prime 2} Z Y,\ 
Z' Z Z' Y,\ Z Z^{\prime 2} Y \ \mbox{and} \ Z^{\prime 3} Y\]
to $\dot Y$, each with factors
\[Tr(Z^2), \quad Tr(Z Z') \quad \mbox{or} \quad Tr(Z^{\prime 2}),\]
  when summed over internal lines.
It follows that, in general, $\dot Z$ will be a linear combination of terms of the form
\beq
{\cal G}_{\mathbf p,\mathbf q}  
\{\Pi_{l=1}^L Tr(Z^{p_{l,1}} Z^{\prime p_{l,2}}  Z^{p_{l,3}} \cdots Z^{\prime p_{l,{r_l}}})\} 
 Z^{q_1} Z^{\prime q_2} Z^{q_3} \cdots Z^{\prime q_{n-1}} Z^{q_n},
\label{eq:GZ}\eeq
where $L$ is the number of closed quark loops in any given diagram.\footnote{We allow for $q_1=0$ if the matrix string starts with $Z'$ on the left, and $q_n=0$ if the string ends with $Z'$ on the right.}
Here 
$\sum_{i=1}^{r_l} p_{l,i}$ is twice the number of Yukawa vertices in loop $l$,
$1+\sum_{l=1}^L \sum _{i=1}^{r_l} p_{l,i}  + \sum_{i=1}^n q_i$ is the twice total number of Yukawa vertices in the diagram\footnote{The $1$ is because the 0th-order diagram, with $\mathbf p = \mathbf q =0$, has one Yukawa vertex.};
${\cal G}_{\mathbf p,\mathbf q}$  is a  scalar in terms of family indices,
but a polynomial in gauge and Higgs couplings and polynomial in sines and cosines of the CKM angles and phase.
                 
Similarly $\dot Z'$ will be a linear combination of terms of the form
\beq
{\cal G}'_{\mathbf p,\mathbf q}  
\{\Pi_{l=1}^L Tr(Z^{\prime p_{l,1}} Z^{p_{l,2}}  Z^{\prime p_{l,3}} \cdots Z^{p_{l,{r_l}}})\} 
 Z^{\prime q_1} Z^{q_2} Z^{\prime q_3} \cdots Z^{q_{n-1}} Z^{\prime q_n},
\label{eq:GZ'}\eeq  

Including leptons will introduce internal loops that are traced over: this can only make ${\cal G}_{k_1,\ldots,k_n}(\alpha_i,\lambda,y_a,y'_a,y'_{l,a},\theta_i,\delta)$ depend on leptonic Yukawa couplings as well, it will not affect the matrix structure of \eqref{eq:GZ}.

Now
\begin{alignat*}{3}
  Z^2  &= U \Lambda^2 U^\dagger= U'(V^\dagger  \Lambda^2 V)U^{\prime \dagger}, && \qquad
Z^{\prime 2} &&= U' \Lambda^{\prime 2}  U^{\prime \dagger}= U (V \Lambda^{\prime 2} V^\dagger) U^\dagger,\\
Z Z'  &= U (\Lambda V \Lambda' V^\dagger) U^\dagger, && \qquad
Z'Z  &&=U (V \Lambda' V^\dagger \Lambda)U^\dagger,\\
       &= U' (V^ \dagger \Lambda V \Lambda') U^{\prime\dagger}, && \qquad
             &&=U' (\Lambda' V^\dagger \Lambda V )U^{\prime \dagger},
\end{alignat*}
and  at all six 1-loop fixed points $V_\alpha^*$, with $\alpha=1,\ldots, 6$,
\bea D^{(*)}_{q_i,\alpha} &=V_\alpha^{* \dagger} \Lambda^{q_i} V^*_\alpha,\\
D^{\prime (*)}_{q_i,\alpha} &= V_\alpha^* \Lambda^{\prime q_i} V_\alpha^{* \dagger}\eea
 are all diagonal matrices --- $V_\alpha^*$ merely permutes the diagonal entries of $\Lambda$ and $\Lambda'$ (the notation $^{(*)}$ is chosen to emphasis that, while $V^*_\alpha$ are fixed points of the CKM running, the Yukawa couplings can still run).
From \eqref{eq:GZ} and  \eqref{eq:GZ'} we conclude that, at the six CKM fixed points,
\beq\dot Z^{(*)} = U D^{(*)} U^\dagger \quad \mbox{and} \quad \dot Z^{\prime (*)} = U' D^{\prime (*)}U^{\prime \dagger}\label{eq:ZUDU}\eeq
with $D^{(*)}$ and $D^{\prime (*)}$ diagonal matrices in family space.

Now defining
\[ A = i U^\dagger \dot U,\qquad A' = i U^{\prime \dagger} \dot U'\] 
we have, in general,
\bea
\dot Z &= U (\dot \Lambda -i[A,\Lambda])U^\dagger, \\ 
\dot Z' &=  U'( \dot\Lambda' -i[A',\Lambda']) U^{\prime \dagger}\eea
and, at the six fixed points \eqref{eq:S_3} of the CKM matrix,
\begin{alignat*}{5} 
\dot Z^{(*)} &= U (\dot \Lambda -i[A^{(*)},\Lambda])U^\dagger  &&= U D^{(*)} U^\dagger
           && \quad \Rightarrow \quad &&
\dot \Lambda -D^{(*)} &&= i[A^{\prime (*)},\Lambda], \\ 
  \dot Z^{\prime (*)}&=  U'( \dot\Lambda' -i[A^{\prime (*)},\Lambda')U^{\prime \dagger} &&= U' D^{\prime (*)} U^{\prime \dagger}
            &&\quad \Rightarrow \quad &&
\dot \Lambda' -D^{\prime (*)} &&= i[A^{\prime (*)},\Lambda'].\end{alignat*}

Since $\dot \Lambda$, $\dot \Lambda'$ are diagonal by definition, and  $D^{(*)}$ 
and $D^{\prime (*)}$ are diagonal at the fixed points of the CKM matrix, we conclude that the commutators $[A^{(*)},\Lambda]$ and $[A^{\prime (*)},\Lambda']$ are diagonal  as well:  they lie in the CSA of $SU(3)$.  Since $\Lambda$ and $\Lambda'$ are already diagonal, this is only possible if $A^{(*)}$ and $A^{\prime (*)}$ are also in the Cartan sub-algebra: thus the commutators  $[A^{(*)},\Lambda]$ and $[A^{\prime (*)},\Lambda']$ vanish  and
\[ \dot \Lambda=D^{(*)}, \qquad \dot\Lambda'=D^{\prime (*)}.\]
Now, from \eqref{eq:CKM}, in general the CKM matrix runs as
\[ \dot V = i (A V - V A')\]
with $A$ and $A'$ in the Lie algebra of $SU(3)$,
hence at the fixed points of the CKM matrix 
\[ A^{(*)} = A^{(*)}_{3} \lambda_ 3 + A^{(*)}_{8} \lambda_8, \qquad 
A^{\prime (*)}= A^{\prime (*)}_{3} \lambda_ 3 + A^{\prime (*)}_{8} \lambda_8,\] 
with $\lambda_3$ and $\lambda_8$ Gell-Mann matrices, are in the CSA to all orders in perturbation theory for any $D^{(*)}$ and $D^{\prime (*)}$.

However $A_3$, $A_3'$, $A_8$ and $A_8^'$ are not physical, they depend on the fermion phases and can be changed by momentum dependent phase changes of the quarks --- in this sense they are gauge potentials for a one-dimensional gauge theory.  
It was shown in \cite{CKM-running}  that, with the parameterisation \eqref{eq:Vdelta},     \hbox{$A^{(*)} V^* - V^* A^{\prime (*)} $} vanishes at 1-loop at the CKM fixed points, but they might not vanish at higher loops.
Consider for example  $(\theta_1,\theta_2,\theta_3,\delta)=(0,0,\frac{\pi}{2},0)$, where
\beq V^*=\bpm 0 & 1 & 0 \\ -1 & 0 & 0 \\ 0 & 0 & 1 \epm \label{eq:V*}\eeq
and
    {\scriptsize{
\[
  i(A^{(*)} V^* \kern -3pt - V^* A^{\prime (*)})  = 
                                                                                \begingroup  
                                         \setlength\arraycolsep{-4pt}
                                         i \bpm 0&  A^{(*)}_3\kern -3pt + \kern -2pt  A_3^{\prime (*)}
                                                   \kern -3pt  +\kern -2pt  \frac{1}{\sqrt 3} (A^{(*)}_8\kern -3pt  -\kern -2pt  A^{\prime (*)}_8) & 0 \\
  A^{(*)}_3\kern -3pt  + \kern -2pt   A_3^{\prime (*)} \kern -3pt  -\kern -2pt  \frac{1}{\sqrt 3} (A^{(*)}_8\kern -3pt  -\kern -2pt   A^{\prime (*)}_8) & 0 & 0 \\
0 & 0 & -\frac{2}{\sqrt 3} (A^{(*)}_8 \kern -3pt  - \kern -2pt  A^{\prime (*)}_8)\epm.
\endgroup
        \]
        }}
The entries in $V^*$ that vanish in \eqref{eq:V*} remain zero under RG evolution,
but the non-vanishing entries may change phase.  But any such change in phase is unphysical, it can be absorbed into the phases of the quarks ---  physically $V^*$ is a fixed point of the RG evolution to all orders in perturbation theory, regardless of the quark phases and the gauge choice for $A_3^*$, $A_3^{\prime *}$, $A_8^*$ and $A_8^{\prime *}$.  The same argument applies to all six fixed points in \eqref{eq:S_3}:  the physics of $V_\alpha^*$ does not  change under RG evolution,  although quark phases might.\footnote{At any one fixed point four quark phase can be chosen so that $A^{(*)}$ and $A^{\prime (*)}$ vanish at that fixed point, but they might not vanish at the other five fixed points in this gauge.
At 1-loop  the quark phases can be chosen so that $A^{(*)}$ and $A^{\prime (*)}$ do all vanish at all six fixed points, but this has not been checked at higher loops.  We conjecture that, even at higher loops, with the  phases chosen so that  $A^{(*)}$ and $A^{\prime (*)}$ vanish at one fixed point then they will vanish at all six, due to $S_3$ symmetry, but this has not verified this explicitly.  }

This completes the proof of the claim in the introduction that the massless running  of the CKM matrix has six fixed points at all orders in perturbation theory, corresponding to the elements of the symmetric group $S_3$, \eqref{eq:S_3}, independently of the values of the Yukawa couplings, which can still run when $V$ is RG fixed. 

\newpage 

\appendix

\end{document}